\documentstyle[12pt]{article}

\begin{document}

\baselineskip=18.6pt plus 0.2pt minus 0.1pt

\makeatletter
\@addtoreset{equation}{section}
\renewcommand{\theequation}{\thesection.\arabic{equation}}

\begin{titlepage}
\title{
\hfill\parbox{4cm}
{\normalsize UFR-HEP/ 2000-07}\\
\vspace{1cm}
   On Hyperkahler Singularities 
  }
\author{
A. {\sc Belhaj}\thanks{{\tt b-adil@emi.ac.ma}}
{} and
 E. H. {\sc  Saidi}\thanks{{\tt  H-saidi@fsr.ac.ma}}
\\[7pt]
{\it  High Energy Physics Laboratory,  Faculty of sciences ,  Rabat ,  Morocco}
}

\maketitle
\thispagestyle{empty}

\begin{abstract}
Using a geometric realization of the $SU(2)_R$ symmetry and a procedure of factorisation of the gauge and $SU(2)_R$ charges, we study the small instanton singularities of the Higgs branch of 
supersymmetric $U(1)^r$ gauge theories with eight supercharges. We derive new solutions for the moduli space of vacua preserving manifestly the eight supercharges. In particular, we obtain an extension of the ordinary ADE singularities for hyperKahler manifolds and show that the classical moduli space of vacua is in general given by cotangent bundles of compact weighted projective spaces.  

\end{abstract}
\end{titlepage}
\newpage
\section{Introduction}

\qquad Over the few past years, there has been an increasing 
interest in studying the moduli space of vacua of the Coulomb 
and Higgs branches of supersymmeric gauge theories with eight 
supercharges in various dimensions. This interest is mostly 
due to the fact that supersymmetry severely restricts 
the quantum corrections to the moduli space metric and allows 
to make many exact computations \cite{SW} . A large class of 
these gauge theories can be realized by  D-brane configurations 
in type II strings on Calabi Yau manifolds \cite{{KKV},{KMV},{BES},{BFS}}
 .\\ 
Recently, a special interest has been  given to the 
analysis of the hypermultiplet gauge invariant moduli 
space near the Higgs branch singularity. This analysis 
has been shown to be relevant for the study of many 
aspects in supersymmetric gauge theories with eight 
supercharges amongst which we quote: (1) the understanding of 
the assymptotic regions of the infrared IR low energy 
limits of the  $N=(4,4)$ supersymmetric gauge theories in 
two dimensions especially within the so called throats of 
the Coulomb and  Higgs branches where the theories are 
typically described by $2d$ $N=4$ conformal Liouville 
theories  \cite{ {SeiW},{AB}}. (2) the study of vector and 
hypermultiplet moduli spaces in the context of
strings compactifications to four dimensions where the low 
energy supergravity has scalar fields in both vector and matter multiplets \cite{W} . (3) the description of stringy instanton moduli space 
as hyperkahler deformations of the classical instanton moduli space with non zero NS-NS B field \cite {{SeiWit},{NS},{N}}. (4) the rederivation  of $2d$ CFT's  from compactifications of superstrings on Calabi Yau fourfolds with a $sp(1)$ hyperKahler singularity \cite{{GVW}}.

The aim of this paper is to study new aspects of hyperkahler singularities of the Higgs branch of supersymmeric gauge theories with eight 
supercharges by using the linear sigma model method. Actually this study extends the analysis of the ADE singularities of Kahler manifolds to hyperKahler ones. It also generalizes known results, on the Coulomb branch of supersymetric gauge theories with four supercharges, to the Higgs branch of supersymetric gauge theories with eight supercharges.\\  
The presentation of this paper is as follows: In section 2, we review  hyperKahler singularities and describe brefly the standard way used in handling the D-flateness eq of supersymmetric theories with eight supercharges.
In section 3, we use the harmonic superspace method of supersymmeric gauge theories with eight supercharges to study the problem of hyperKahler singularities. This way of doing has many advantagesas  first it preserves manifestly the eight supercharges secon it keeps all the three FI parameters non zero without need to make any appropriate choice.   Our method , which is based on realizing the $ SU(2)_R$ invariance a by helf harmonic analysis on $ S^2$,  permits to exhibit explicitly the role of the three Kahler structures of the gauge invariant hyperkahler moduli space.  In Section 4, we study the solving of the D-flateness eqs by introducing a method based on the factorization of the charges of the gauge and the $SU(2)_R$ symmetries. We give two classes of solutions of these eqs preserving manifestly the eight supercharges and recverring known resultts on ADE singularities  by particular breaking half of the supersymetry. In section 5, we give our conclusion.

\section{Generalities on Hyperkahler singularities }

\qquad We begin by noting that Calabi Yau fourfolds can develop singularities of many types; this includes the 
$C^4\over{Z_4}$ orbifold, the ADE hypersurfaces considered recently 
in \cite{GVW} and the so called hyperkahler 
singularity we are intersted in here. We review hereafter the example of the hyperkahler cotangent bundle  $T^*(CP^2)$ of complex projective space  $(CP^2)$ and its generalisation  $T^*(CP^n)$ for $ n>2$. To that purpose, consider $2d $ $ N =4$ supersymmetric $U(1)$ gauge theory with one FI isotriplet 
 coupling $\vec{\xi}=(\xi^1,\xi^2,\xi^3)$ and three hypermultiplets of charges $q^i_a=q^i=1;  i=1,2,3$. The moduli space of this 
gauge model are obtained by solving 
\begin{equation}
\label{6}
\sum_{i=1}^{3}[ \overline{\varphi}_{i\alpha} {\varphi}_i^\beta+
\varphi_{i\alpha}\overline{\varphi}_i^\beta]=\vec{\xi}\vec{\sigma}_\alpha^\beta,
\end{equation} 
which by the way is just a special situation of the following eqs: 
\begin{equation}
\sum_{i=1}^{n} q_a^i[\varphi_i^\alpha \overline{\varphi}_{i\beta}+
\varphi_{i\beta}\overline{\varphi}_i^\alpha]=
\vec{\xi_a} \vec{\sigma}^\alpha_\beta\qquad;a=1,...,r.
\end{equation}

For later use it is interesting to note that eqs(2.2) have a formal analogy with the sigma model vaccum eqs of $2d$ $ N=2 $  supersymmetric $U(1)^r$ gauge  theory involved in the 
analysis of the Coulomb branch of IIA superstrings on Calabi Yau threefolds with ADE 
singularities.
\begin{equation}
\sum_i q_a^i|X_i|^2=R_a; \qquad a=1,...,r. 
\end{equation}
In eqs (2.3), the $X_i$'s are complex scalar fields, the $R_a$'s are FI 
couplings and the $q_a^i$'s are the $U(1)^r$ gauge charges of the $X_i$'s which, 
for reference, read in the case of a $SU(n)$ singularity as :
\begin{equation}
\ q_a^i=-2\delta_a^i+\delta_a^{i-1}+\delta_a^{i+1},
\end{equation}
with the remarkable equality
\begin{equation}
\label{4}
\sum_i q_a^i=0.
\end{equation} 
\\ Eqs (2.1) is a system of three eqs which, up to replacing 
the Pauli matrices by their expressions and using the $SU(2)_R$ transformations 
$\varphi^{\alpha}=\varepsilon^{\alpha\beta}\varphi_ {\beta}$ with 
$\varepsilon_{12}=\varepsilon^{21}=1$ and 
$\overline{(\varphi^\alpha)}=\overline{\varphi}_\alpha$, split as follows:

\begin{equation}
\begin{array}{lcr}
\sum\limits_{i=1}^3( |\varphi^1_i|^2-|\varphi^2_i|^2) &= \xi^3 \qquad &
(a)\\ 
\sum\limits_{i=1}^3 \varphi^1_i \overline{\varphi}_{i2}&=\xi^1+i{\xi^2}\qquad 
&(b)
\\ 
\sum\limits_{i=1}^3 \varphi^2_i \overline{\varphi}_{i1}&=\xi^1-i{\xi^2}\qquad 
&(c).
\end{array}
\end{equation}
The moduli space of zero energy states of the classical gauge theory is the space 
of the solutions of eqs (2.1) and (2.6) divided by the action of the $U(1)$ gauge group. The solutions of eqs (2.1) depend on the values of the FI couplings. For the case where 
$\xi^1=\xi^2=\xi^3=0$, the moduli space has an $SU(3)\times SU(2)_R$ symmetry; it 
is a cone over a seven manifold described by the eqs:    
\begin{equation}
\sum_{i=1}^3(\varphi_{\alpha i}\overline{\varphi}_i^\beta-
\varphi_i^\beta\overline{\varphi}_{\alpha i})= \delta{ _\alpha ^\beta}.
\end{equation}
 For the case $\vec{\xi}\neq\vec{0}$, the abovementioned 
$SU(3)\times SU(2)_R$ symmetry is explicitly broken down to 
$SU(3)\times U(1)_R$. In the remarkable case where  $\xi^1=\xi^2=0$  and 
$\xi^3 >0$, it is not difficult to see that eqs (2.1) 
describe the cotangent bundle of $CP^2$. Indeed making the change
\begin{equation}
\label{9}
{\psi_i}={\varphi_{i}^{1}\over{[\sum\limits_{i=1}^3|\overline{\varphi}_{i2}|^2+
\xi^3]^{1\over2}}},
\end{equation}
and putting back into eq (2.6.a), one discovers that $\psi_{i}$'s 
satisfy $\sum \limits_ {i}{|\psi_i|^2}=1$. The $\psi_i$'s parametrize the 
$CP^2$ space. On the other hand with $\xi^1=\xi^2=0$  
 conditions, eqs (2.6-b,c) may be interpreted to mean that 
$ \overline{\varphi}_{2i}$ lies in the cotangent space to $CP^2$
 at the point determined by $\psi_i$. Note that though we are usually 
allowed to make the choice $\xi^1= \xi^2=0$ by using an appropriate 
$ SU(2)_R$ transformation, we shall consider in section 4,  
the generic cases where $\xi^1, \xi^2$ and $\xi^3$ are all of them non 
zero. For the time being let us note that the previous analysis may 
be extended to the cases of $2d$ $ N=4$ supersymmetric $U(1)$ gauge linear 
sigma model involving $n+1$ hypermutiplets with charges $q^i=1;  i=1,..., n+1 $ 
and transforming in the fundamental representation of $SU(n+1)$. The vaccum energy 
equations of this $U(1)$ gauge model read as :
\begin{equation}
\begin{array}{lcr}
\sum\limits_{j=1}^{n+1}( |\varphi_{j}^{1}|^2-|\overline{\varphi}_{2j}|^2)&=\xi^3\\
\sum\limits_{j=1}^{n+1}( \varphi_{j}^{1}\overline{\varphi}_{2j})&=\xi^1+i\xi^2.
\end{array}
\end{equation}
For $\xi^1= \xi^2=0$ and $\xi^3  >0$, the classical moduli 
space of the gauge theory is given by the cotangent bundle of 
complex $n$ projective space: $T^*CP^{n}$ . For $\xi^1= \xi^2=\xi^3=0$, 
one has just the conifold  singularity of $n$ dimensional complex manifolds.
 
 \section{Hyperkahler singularities in Harmonic superspace}
\qquad Eqs(2.2) is a system of $3r$ equations which shares some general features with the usual $2d$ $  N=2$ supersymmetic D-flatness conditions eqs(2.3). To solve them, we shall use a different method inspired from harmonic superspace. The latter was shown to be the appropriate space to deal with supersymmetric quantum field theories with eight supercharges. We shall not develop here the basis of this superspace; details may be found in  \cite{{LS},{SS}}.  Our approach will be done in two steps: First we use $SU(2)_R$ harmonic analysis allowing us to realise  $SU(2)_R$ representations as special functions on $ S_R^2$. The index R 
carried by $S_R^2$ and $U(1)_R$  refers to the $ SU(2)_R$. Second, we introduce a convenient change of 
variables based on factorizing the $U(1)^r_G$ gauge charges and the  $SU(2)_R$ ones. 
We begin by describing the first step of this program. Introducing the following  $2\times 2$ matrix       
\begin{equation}
\label{11}
U=\left(\matrix{
u^+_1&u^+_2\cr
u^-_1&u^-_2
\cr}\right)
\end{equation}
and solving the isospin $1\over{2}$ $SU(2)_R$ representation constraints 
namely the unimodularity $det U= 1$
and the unitarity $U^+U=U^+U=I$  conditions, one discovers the defining eq of the unit  $S_R^3$ sphere:

\begin{equation}
\label{12}
\begin{array}{lcr}
u^{{\pm}{\alpha}}=\epsilon^{{\alpha}{\beta}}u^{\pm}_{\beta}; \qquad 
\overline{u}^{+\alpha}=u^-_{\alpha}; \qquad \epsilon_{\alpha\beta}=-
\epsilon_{\beta\alpha}\\

u^{+\alpha}u^{-}_{\alpha}=1,\qquad u^{+\alpha}u^{+}_\alpha =u^{-\alpha}u^{-}_\alpha=0.
\end{array}
\end{equation}
 
Moreover, using  the $u_\alpha^\pm$ variables,  
the $SU(2)_R$ algebra is realized as differential operators on the space 
of harmonic functions on $S_R^3$:
\begin{equation}
\begin{array}{lcr}
D^{++}=u^{+ \alpha}{\partial\over{\partial u^{-\alpha}}};\qquad D^{--}=
u^{- \alpha}{\partial\over{\partial u^{+\alpha}}}\\
2D^{++}=[D^0,D^{++}];\qquad -2D^{--}=[D^0,D^{--}]\\
D^0=[D^{++},D^{--}]=u^{+ \alpha}{\partial\over{\partial u^{+\alpha}}}-
u^{- \alpha}{\partial\over{\partial u^{-\alpha}}}
\end{array}
\end{equation}
To study the $SU(2)_R$ representations by using the harmonic variables, 
it is more convenient to consider harmonic functions $ F^q(u_\alpha^\pm)$  
with definte $U(1)_R$ charge q ; that is functions $ F^q(u_\alpha^\pm)$  
satisfying the eigenfunction eq 
\begin{equation}
[D^0,F^q]=qF^q.
\end{equation}
These functions  $F^q$ have a global harmonic expansion of total charge $q$ 
and carry $SU(2)_R$ representations. For example, taking $q=2$ and choosing  
$F^{++}$ as: 
\begin{equation}
\label{16}
F^{++}(u^{\pm}_\alpha)=u_{(\alpha}^+u_{\beta)}^+F^{(\alpha\beta)},
\end{equation}
one sees that $F^{++}$ is the highest state of the isovector representation of 
$SU(2)_R$;i.e: 

 \begin{equation}
[D^0,F^q]=qF^q, \qquad [D^{++},F^q]=0.
\end{equation}

Note that one can usually realize $F^{++} $ as bilinears of 
isospinors $f^+$ and $\overline {f}^+$ as  follows:
\begin{equation}
\begin{array}{lcr}
F^{++}=if^+\overline{f}^+=iu_{(\alpha}^+u_{\beta)}^+f^{(\alpha}\overline{f}^{\beta)}.\\
\end{array}
\end{equation} 
The complex number  $i$ in front of the the factor of the right hand 
side of the above eqs ensures the reality condition of the isotriplet 
representation. After this digression on the $SU(2)_R$ harmonic analysis, 
we turn now to eqs (2.2) which up on multiplying its both sides by $u_{(\alpha}^+u_{\beta)}^+$, we get:
\begin{equation}
\sum\limits_{j}q^j_a \varphi_{j}^{+}\overline{\varphi}_{j}^{+}=
-i\xi^{++}_a.
\end{equation} Eqs (3.8)  are also 
the D-flatness eqs one gets if one is using the $2d$ $ N=(4,4)$  
harmonic superspace formulation of $2d$ $N=4$ gauge theories \cite{{LS},{SS}}.

In the end of this section, we would like to make two comments: First one can use the isospinor bilinear realization 
of isotriplets eqs (3.7) to represent the Kahler parameters $\xi_a^{++}$ as follows:

\begin{equation}
\begin{array}{lcr}
\xi_a^{++}=i\zeta_a^+\overline{\zeta}^+_a=iu_{(\alpha}^+u_{\beta)}^+
\zeta^{(\alpha}_a{\overline{\zeta}^{\beta)}_a}\\
\zeta_a^{\pm}=u_{\alpha}^{\pm}\zeta^{\alpha}_a;\qquad\overline\zeta^{\pm}_a=
u_{\alpha}^{\pm} \overline{\zeta}^{\alpha}_a,
\end{array}
\end{equation}  
where $\zeta_a^\alpha$ and  $ \overline\zeta_a^\alpha$ may, roughly speaking, 
be viewed as  the square roots of the FI couplings $\xi_a^{(\alpha\beta)}$. 
Similar  relations involving $\zeta_a^{\pm}$ and  $ \overline{\zeta}_a^{\pm}$ 
for $\xi_a^0$ and  $ \xi_a^{--}$ may be also written down. Putting back these 
relations in eqs (3.8), one gets
\begin{equation}
\sum_{j}q^j_a \varphi_{j}^{+}\overline{\varphi}_{j}^{+}=\zeta_a^+\overline{\zeta}^+_a
=u_{(\alpha}^+u_{\beta)}^+\zeta^{(\alpha}_a\overline{\zeta}^{\beta)}_a.
\end{equation}
Second, simplify further eqs (3.10) by making 
an extra change of variables which turns out to convenient when discussing the moduli 
space of gauge invariant vacua of eqs (2.2). This extra change consists to use the 
mapping $ R^3=R^+ \times S^2 $ to write the FI isovectors $\xi_a^{++}$ as
\begin{equation}
\label{24}
\xi_a^{++}= R_a\eta^+_a \overline{\eta}^+_a=r_a^2\eta^+_a \overline{\eta}^+_a,
\end{equation}
or equivalently by using the isospinors $\zeta_a^\alpha $ and  
$ \overline{\zeta}_a^{\alpha}$ introduced previously: 
\begin{equation}
\label{25}
\begin{array}{lcr}
\zeta^{\pm}_a&=u^{\pm}_\alpha \zeta^\alpha_a; \qquad \overline{\zeta}^{\pm}_a
=u^{\pm}_\alpha \overline{\zeta}^\alpha_a\qquad &(a)\\
\zeta^\alpha_a&=r_a\eta_a^\alpha;\qquad \overline{\zeta}^\alpha_a
=r _a\overline{\eta}_a^{\alpha}\qquad &(b)\\
 \zeta^\alpha_a\overline{\zeta}_{a\alpha}&=r^2_a\geq 0 \qquad  
\qquad \qquad \qquad &(c).
\end{array}
\end{equation}
Eqs (3.11) and (3.12) tell us that the $R_a$'s $(R_a=r_a^2\geq 0)$ 
are the radial variables and the  $\eta_a^\alpha $'s  and    
$ \overline {\eta}_{a\alpha} $'s,  which satisfy       
 \begin{equation}
\label{26}
\eta^\alpha_a\overline{\eta}_{a\alpha}=1;\qquad\eta^\alpha_a{\eta}_{a\alpha}= 
\overline{\eta}^\alpha_a\overline{\eta}_{a\alpha}=0,
\end{equation}
parametrize the two spheres $ S^2_a$. The $r_a^2$ and $\eta_a^\alpha $ and 
$ \overline {\eta}_{a\alpha} $  are in one to one correspondance with the 
$r$ FI isovectors. In other words eqs (3.13) describe a collection of $r$ 
unit two spheres which altogether with the $r_a$ conical variables of $(R^3)^r$ 
give the $3r$ parameters of  the $r$ FI isovector couplings.

\section{Solving the D-flateness eqs}

 In this section, we solve eqs(2.2) by developing a factorization method of the gauge and $SU(2)_R$ charges carried by the hyperKahler moduli. This factorization should satisfy the following natural constraints: \\
(a) The splitting of the gauge and $SU(2)_R$ charges preserves the 
eight supercharges of the gauge theory. \\
(b)It recovers the results of \cite{GVW}, extends the ADE models of Kahler backgrounds \cite{KMV}and, up on breaking half of the eight supercharges, gives the standard ADE results.\\ 
(c) It has a geometrical interpretation.\\
The factorization of hyperKahler moduli  solving the above constraints is given by: 

 \begin{equation}
\begin{array}{lcr}
\varphi^+_j=X_j\eta^+_j+\gamma Y_j \overline{\eta}^+_j\\
\overline{\varphi}^+_j=\overline{X}_j \overline{\eta}^+_j-\gamma \overline{Y}_j 
\overline{\eta}^+_j,
\end{array}
\end{equation}
where $\eta^+_j$ and $ \overline{\eta}^+_j$ are as in eqs (3.12); that is:
\begin{equation}
\begin{array}{lcr}
\eta^+_j=u^+_\alpha\eta^\alpha_j;\qquad \overline{\eta}^+_j=u^+_\alpha
\overline{\eta}^\alpha_j \\
\eta^\alpha_j \overline{\eta}_{\alpha j}=1;\qquad \eta^\alpha_j {\eta}_{\alpha j}=
\overline{\eta}^\alpha_j \overline{\eta}_{\alpha j}=0. 
\end{array}
\end{equation} 
In eqs(4.1), $X_j$  and $Y_j$, $j=1,...,n$ are complex fields carrying no $SU(2)_R$ 
charge. The parameter $\gamma$ takes the values $\gamma=0$ or $\gamma=1$  and allow to distinguish two classes of solutions we will give hereafter.   Moreover the 
quantities $X_j$ , $Y_j$  and $\eta^+_j$ and $ \overline{\eta}^+_j$ of 
eqs(4.1) behave under $ U(1)^r$ gauge and $ SU(2)_R$ transformations 
as follows:        
\begin{equation}
\begin{array}{lcr}
U(1)^r : X_j \longrightarrow X^\prime_j=\lambda ^{q^j_a}X_j \\
\qquad Y_j\longrightarrow Y^\prime_j=\lambda ^{q^j_a}y_j \\
\qquad\eta^+_j\longrightarrow  \eta\prime^+_j= \eta^+_j  \\
\qquad \overline {\eta}^+_j\longrightarrow  \overline{\eta}\prime^+_j= 
\overline{\eta}^+_j ;
\end{array}
\end{equation}
and
\begin{equation}
\begin{array}{lcr}
 U(1)_R : X_j \longrightarrow X^\prime_j=X_j\\
\qquad Y_j\longrightarrow Y^\prime_j=Y_j\\
 \qquad\eta^+_j\longrightarrow  \eta\prime^+_j= e^{i\theta}\eta^+_j\\ 
\qquad \overline {\eta}^+_j\longrightarrow  \overline{\eta}\prime^+_j= 
e^{i\theta}\overline{\eta}^+_j .
\end{array}
\end{equation} 
Actually eqs (4.3-4) define the factorization of the gauge charges and 
$ SU(2)_R$ ones of the vacum moduli. In what follows we shall use the splitting eqs(4.1) to solve 
eqs (2.2) which, by help of the analysis of section 3, may be put in the form:         
\begin{equation}
\sum_{j}q^j_a \varphi_{j}^{+}\overline{\varphi}_{j}^{+}=r^2_a\eta_a^+
\overline{\eta}^+_a.
\end{equation}
We give herebelow two classes of solutions of these eqs. The first class describes 
a generalisation of the usual ALE surfaces with ADE singularities. The second class leads to new models which flow in the infrared to $2d$ $ N=(4,4)$ conformal field theories. 

\subsection{ Generalized ADE hypersurfaces } 

  Eqs(2.2) looks formally like eqs (2.3); their solutions are then expected to describe generalisations of the standard eqs of ADE singularities associated with $2d$ $N=2$ linear sigma models. Imposing the ADE Calabi Yau condition 
\begin{equation}
\label{32}
\begin{array}{lcr}
\sum\limits_{j}q^j_a=0,\qquad a=1,...,n-1,
\end{array}
\end{equation}
one can imitate the analysis of $ N=2$ linear $ \sigma$ models 
and build the gauge invariant moduli in terms of the $\varphi_{j}^{+}$  fields. In the $SU(n)$ case for instance where $q_a^j$ is given by eq(2.4), there are three gauge invariant moduli; $U^{+{n(n+1)\over2}}$, $V^{+{n(n+1)\over2}}$  
and $Z^{+{(n+1)}}$   carrying ${n(n+1)\over2}$ , ${n(n+1)\over2}$ and 
${(n+1)}$ $U(1)_R$  Cartan charges respectively. These invariant read as:      
\begin{equation}
\label{33}
\begin{array}{lcr}
U^{+{n(n+1)\over2}}=\prod\limits_{j=0}^n(\varphi^+_j)^{n-j}\\

V^{+{n(n+1)\over2}}=\prod\limits_{j=0}^n(\varphi^+_j)^j\\

Z^{+{(n+1)}}=\prod\limits_{j=0}^n(\varphi^+_j).
\end{array}
\end{equation}
They satisfy the following remarkable equation,
\begin{equation}
\label{34}
U^{+{n(n+1)\over2}}V^{+{n(n+1)\over2}}=[Z^{+{(n+1)}}]^n.
\end{equation}
Eq (4.8) generalizes the usual equation of the ALE surface with $SU(n)$ singularity:    
\begin{equation}
\label{35}
uv=z^n.
\end{equation}
 To better see the structure of eq (4.8), we use the $\varphi^+_j$'s moduli factorization described earlier. Taking  $\gamma=0$, 
the general splitting eqs (4.1) reduces to: 
\begin{equation}
\label{36}
\varphi^+_j=x_j\eta^+_j ; \qquad \overline{\varphi}^+_j= \overline{x}_j 
\overline{\eta}^+_j,
\end{equation}
where $X_j$ and $\eta^+_j$ behave under gauge and $SU(2)_R$ transformations 
as in eqs (4.3-4). Note that like $\varphi_j^{\pm}$, the 
realization $X _j \eta_j^{\pm}$ carries for each value of j, four real 
degrees of freedom; two coming from $ X_j$ and the two others from  
the parameters the sphere $ S^2$ described by $\eta_j^{\pm}$. Under $ 2d $ $N=4$ supersymmetric transformations which  may be conveniently  
expressed as $4d$ $N=2$ supersymmetric transformations of fermionic parameters  
 $\epsilon ^{\pm}$  and $ \overline{\epsilon} ^{\pm}$, we have:
\begin{equation}
\label{38}
\delta\varphi^+_j=\epsilon ^+\psi_j+\overline{\epsilon} ^+ \overline{\chi}_j;
\end{equation}
where $\psi_j$ and $\overline{\chi}_j$ are  the Fermi partners of the 
$\varphi_j^{\pm}$ scalars. $(\varphi^{\pm}_j,\psi_j,\overline{\chi}_j)$ 
constitute altogether the $ 4d$ $N=2$ free hypermultiplets. Using the 
splitting principle by factorizing  $ \epsilon^+$ as $\epsilon\eta^+$  
and $\overline{\epsilon}^+={\overline{\epsilon} } {\overline{\eta}^+}$, 
and using eqs (4.10-11), we get
\begin{equation}
\label{39}
\eta^+_j\delta X_j+X_j\delta\eta^+_j =\overline{\eta}^+\epsilon 
\psi_j+\overline{\eta}^+{\overline{\epsilon }} \overline{\chi}_j,
\end{equation}
or equivalenty
\begin{equation}
\label{38}
\eta^\alpha_j\delta X_j+X_j\delta\eta^\alpha_j =\overline{\eta}^\alpha\epsilon 
\psi_j+\overline{\eta}^\alpha\overline{\epsilon } \overline{\chi}_j.
\end{equation}
    Putting eqs (4.10) back into eqs (4.5), we obtain 
\begin{equation}
\sum_{j}q^j_a |X_{j}|^2\eta^+_j\overline{\eta}^+_j=r^2_a\eta_a^+\overline{\eta}^+_a
\end{equation}
and   
\begin{equation}
\begin{array}{lcr}
U^{+{n(n+1)\over2}}=uM^{+{n(n+1)\over2}}\\
V^{+{n(n+1)\over2}}=vN^{+{n(n+1)\over2}}\\
Z^{+{(n+1)}}=zS^{+{(n+1)}} ;
\end{array}
\end{equation}
where $ u, v, z$ and $M^{+{n(n+1)\over2}}$, $ N^{+{n(n+1)\over2}}$ and $ S^{+{(n+1)}}$ 
are gauge invariants given by:
\begin{equation}
\begin{array}{lcr}
u=\prod\limits_{j=0}^n X_j^j&;\qquad N^{+{n(n+1)\over2}}&=
\prod\limits_{j=0}^n(\eta^+_j)^{n-j}\\
v=\prod\limits_{j=0}^n X_j^{n-j}&;\qquad M^{+{n(n+1)\over2}}&=
\prod\limits_{j=0}^n(\eta^+_j)^j\\
z=\prod\limits_{j=0}^n X_j&;\qquad S^{+{(n+1)}}&=\prod\limits_{j=0}^n\eta^+_j.
\end{array}
\end{equation}  
Note that  $ u, v$ and $z$ verify the relation (4.9) and $ M^{{+{n(n+1)\over2}}}$ , 
$ N^{{+{n(n+1)\over2}}}$ and $ S^{+{(n+1)}}$
satisfy eq (4.8). Note moreover that eqs (4.14-15) may be brought to more familiar forms if one requires that all 2-spheres are identified; i.e:\\
(i) The $(n+1)$ 2- spheres parametrized by  the $\eta_j$'s .\\
(ii) The $(n-1)$ $\eta_a$ 2-spheres used in the parametrization of the FI couplings 
eqs (3.13). \\
(iii) The $ \eta^+ $ 2-sphere involved in the factorization of the supersymmetric 
parameter $\epsilon^+$ eq(4.12).\\
 Put differently, we require the following constraint eq to hold: 
\begin{equation}
\eta^+_j=\eta^+_a=\eta^+.
\end{equation}
With this identification, eqs (4.14) reduce to the well known 
D- flatness conditions of the $ U(1)^r$ supersymmetric gauge theory with four supercharges; namely:
\begin{equation}
\sum_{j}q^j_a |X_{j}|^2=r^2_a.
\end{equation}
Moreover eq (4.8) reduces to the usual ALE surface with $SU(n)$ singularity 
(4.9) since the
 $M^{+{n(n+1)\over2}}$, $ N^{+{n(n+1)\over2}}$ and $ S^{+{(n+1)}}$ gauge 
invariant become trivial; they are given by powers of $\eta^+$ as shown here below.
 \begin{equation}
\begin{array}{lcr}
M^{+{n(n+1)\over2}}=(\eta^+)^{{n(n+1)\over2}}=N^{+{n(n+1)\over2}}\\
S^{+(n+1)}=(\eta^+)^{n+1}.
\end{array}
\end{equation}
In the general case where the gauge charges $q^j_a$ of the $X_j$'s satisfy the constraints(2.5), eqs(4.18) is the vaccum energy of $ 2d$ $ N=2$ supersymmetric 
linear  $\sigma$ models. Thus, the classical moduli space $\cal M$ of the gauge invariant 
vacua of eqs(4.17-18) is given by the product of the 2-sphere parametrized by $\eta^+$; eq(4.17)and the moduli space of the gauge invariant solutions of $2d$ $ N=2$ vaccum energy states. In other words: 
$$ {\cal M}={C^{n+1}\over {{C^*}^{n-1}}}\times S^2.$$
Note that the identification constraint eq(4.17) has a nice interpretation; it  
breaks explicitly half of the eight supersymmetries leaving then four supercharges manifest. These four supercharges are behind the reduction of eqs(4.14) down to eqs(4.18)leading to the standard ADE models. This feature is immediately derived by combining eqs(4.13) and (4.17) as follows :
\begin{equation}
\eta^\alpha\delta X_j+X_j\delta\eta^\alpha =\overline{\eta}^\alpha\epsilon \psi_j+
\overline{\eta}^\alpha\overline{\epsilon } \overline{\chi}_j.
\end{equation} 
Then multiplying both sides of this identity by $\overline{\eta}_\alpha$; one gets, 
after using eqs(4.2): 
\begin{equation}
\label{39}
\delta X_j=\epsilon \psi_j. 
\end{equation}
This eq gives the usual supersymmetric transformations of the complex scalars of the $2d$ $ N=2$ chiral multiplets. This completes our check of consistency of the generalised 
$ SU(n)$ hypersurface singularity (4.8). Before going ahead let us summarize in few  
words what we have been doing. Starting from eqs(2.2), we have shown that it is possible to put them into the equivalent form (4.5). The  corresponding moduli 
space of gauge invaraint vacua is given by eq (4.8) which reduces to the standard 
ALE space with $ A_{n-1}$ singularity up on imposing the factorization eqs (4.10) 
and the conditions (4.17). The latters break four of the original 
eight supercharges. To restore the eight supersymmetries by still using 
the constraint (4.17), we should take $\gamma$ non zero; say $\gamma=1$.  Non 
zero $\gamma$ restores four extra supercharges which add to the old four existing 
ones carried by eq (4.10). Note in passing that a naive analysis of eqs(4.5) suggests that the gauge invariant moduli 
space of vacua  $\cal M$ of eq (4.10) is given, for the generalized $SU(n)$ singularity 
eq (4.8), by  the usual ALE space with $SU(n)$ singularity times the  two-sphere power $2n$ :
$$ {\cal M}={C^{n+1}\over {{C^*}^{n-1}}}\times{( S^2)}^{2n},$$
where  $(n+1)$ two spheres come from  the $ \varphi^+_j$'s as shown in eqs (4.10) and  
$(n-1)$ two spheres come from the FI couplings. \\
     
\subsection {Second Solution}
Putting $\gamma=1$ in eqs(4.1) and (4.5), we get the 
system of three eqs given by:
\begin{equation}
\begin{array}{lcr}
\sum\limits_{j}q^j_a(|X_j|^2-|Y_j|^2)\eta_j^+\overline{\eta}_j^+&=r^2_a\eta_a^+
\overline{\eta}_a^+ \qquad &(a)\\
\sum\limits_{j}q^j_a(X_j\overline{Y}_j)\eta_j^+{\eta}_j^+&=0 \qquad &(b)\\
\sum\limits_{j}q^j_a(\overline {X}_j{Y_j})\overline{\eta}_j^+\overline{\eta}_j^+&=0. 
\qquad &(c)
\end{array}
\end{equation}
At this level no constraint on the FI couplings has been imposed yet. If moreover we require that 
all the two sphere $\eta^+_j$ and $\eta^+_a$ are identified as in eq(4.17); the above system reduces to:
\begin{equation}\label{44}
\begin{array}{lcr}
\sum\limits_{j}q^j_a(|X_j|^2-|Y_j|^2)&=r^2_a \qquad &(a)\\
\sum\limits _{j}q^j_a(X_j\overline{Y}_j)&=0 \qquad &(b)\\
\sum\limits _{j}q^j_a(\overline {X}_j{Y_j})&=0. \qquad &(c)
\end{array}
\end{equation}
Eqs(4.22) have remarkable features which have nice interpretations. Though the 
$ q^j_a$ gauge charges of the hypermultiplet moduli are not required to add to zero as in eq (2.5), eq (4.22.a) behave exactly as the D-flatness condition of $2d $ $N=4$ 
supersymmetric $U(1)^r$ gauge theory. The point is that eqs(4.22.a) involve twice 
the number of fields of eqs(2.3), but with opposite charges $q^j_a$. Put 
differently; eq(4.22) involve two sets of fields ${X_j}$ and ${Y_j}$ of charge 
$q^j_a$ and ($-q^j_a$) respectively. The sum of gauge charges of the $X_j$'s 
and $Y_j$'s add automatically to zero eventhough eq(2.5)is not fulfilled.
\begin{equation}
\sum_{j}q^j_a+\sum_{j}(-q^j_a)=0.
\end{equation}
Eqs(4.24) go beyond the constraint eqs(2.5)and so models with $\gamma=1$ flow in the IR to $2d$ $ N=(4,4)$ superconformal 
models extending the usual $2d$ $ N=(2,2)$ ADE ones. Moreover 
eqs(4.24) may be fulfilled in different ways;
either by taking all charges $q^j_a$ of the $U(1)^r$ gauge theory to be 
positive; say $q^j_a=1; a=1, ...,r; j=1,..., n$, or part of the $q^j_a$'s 
are positive and the remaining ones are negative. In the case of a supersymmetric $ U(1)$  
gauge theory with $(n+1)$
hypermutiplets with gauge charges equal to one, eqs(4.23.a) describe a $ CP^n$ 
manifold while eqs(4.23-b,c) which read as
\begin{equation}
 \sum_{j} X_j\overline {Y_j}=0 ,
\end{equation} 
together with their complex conjugate, show that the  $\overline {Y_j},$'s are in the cotangent space of $ CP^n$ at the point $x_j=X_j/{[\sum\limits_i|Y_i|^2+r_a^2]^{-{1\over2}}}$.
Observe that in case where some of the positive charges $ q^j_a$ of 
$ U(1)^r$ gauge theory are not equal to one, the corresponding moduli space is just the cotangent bundle of some weighted complex projective space, $ T^*(WP^n)$. 
Obesrve moreover that in the infrared limit this gauge theory flows to a $2d$  
$ N=(4,4)$ conformal field theory with central charge $C=6n$.

\section{Conclusion}

 In this paper we have studied the solutions of the D-flatness 
eqs of supersymmetric $ U(1)^r$ gauge theories with eight supercharges by 
using the harmonic superspace linear sigma model approach . We have studied the blown up of hyperkahler singularities and shown that they are given by
cotangent bunbles of compact weighted projective spaces. The latters 
depend on the number of hypermultiplets and gauge supermultiplets involved in the gauge model one is considering. This examination extends the standard linear sigma model analysis performed for 
the Kahler Coulomb branch of supersymmetric gauge theories with four supercharges. 
Our method, which go beyond standard analysis where only half of the eight supersymmetries are apparent, preserves manifesty all the eight supersymmetries and is done in two steps. First we have used a geometric realization 
of the $SU(2)_R$ symmetry to transform the D-flatness eq in a more convenient form quite easy to handle. Second we have introduced a method of factorization of gauge and $SU(2)_R$ charges of the 
hypermultiplet moduli. This factorisation involves an index $\gamma$  
taking the values $\gamma=0$ or $\gamma=1$ which allow to distinguish two classes of 
solutions of eqs(2.2)preserving the eight supercharges. For $\gamma=0$, 
we have obtained a generalisation of the ADE complex surfaces which read in the non affine case as:
\begin{equation}
\label{5}
\begin{array}{lcr}
A_{n-1}:\qquad U^{+{n(n+1)\over2}}V^{+{n(n+1)\over2}}=[Z^{+{(n+1)}}]^n\\
D_{n}:\qquad(x^{++})^{n}+x^{++}(y^{+(n-1)})^2+(z^{+n})^{2}=0\\
 E_6:\qquad (x^{+6})^2+(y^{+4})^3+(z^{+3})^4=0\\
 E_7:\qquad (x^{+9})^2+(y^{+6})^3+y^{+6}(z^{+4})^3=0\\
 E_8:\qquad (x^{+15})^2+(y^{+10})^3+(z^{+6})^5=0.
\end{array}
\end{equation}
 The above eqs reproduce the 
standard ADE singularities by partial breaking of $2d$ $N=4$ supersymmetry down to $2d$ $N=2$. 
For $\gamma=1$, we have found new models which flow in the infrared to $2d$ $N=(4,4)$ scale invariant models with central charge c= 6k. 

 This research work has been supported  by the program PARS number phys 27/ 372/98 CNR.

\end{document}